\documentclass[twocolumn,floatfix,aps,superscriptaddress]{revtex4}
\usepackage{graphicx}
\usepackage{amsmath}
\usepackage{amssymb}
\usepackage{bm}
\usepackage{hyperref}

\begin{document}

\title{Deterministic creation and braiding of chiral edge vortices}
\author{C. W. J. Beenakker}
\affiliation{Instituut-Lorentz, Universiteit Leiden, P.O. Box 9506, 2300 RA Leiden, The Netherlands}
\author{P. Baireuther}
\affiliation{Instituut-Lorentz, Universiteit Leiden, P.O. Box 9506, 2300 RA Leiden, The Netherlands}
\author{Y. Herasymenko}
\affiliation{Instituut-Lorentz, Universiteit Leiden, P.O. Box 9506, 2300 RA Leiden, The Netherlands}
\author{I. Adagideli}
\affiliation{Faculty of Engineering and Natural Sciences, Sabanci University, Orhanli-Tuzla, Istanbul, Turkey}
\author{Lin Wang}
\affiliation{Kavli Institute of Nanoscience, Delft University of Technology, P.O. Box 5046, 2600 GA Delft, The Netherlands}
\author{A. R. Akhmerov}
\affiliation{Kavli Institute of Nanoscience, Delft University of Technology, P.O. Box 5046, 2600 GA Delft, The Netherlands}
\date{September 2018}

\begin{abstract}
Majorana zero-modes in a superconductor are midgap states localized in the core of a vortex or bound to the end of a nanowire. They are anyons with non-Abelian braiding statistics, but when they are immobile one cannot demonstrate this by exchanging them in real space and indirect methods are needed. As a real-space alternative, we propose to use the chiral motion along the boundary of the superconductor to braid a mobile vortex in the edge channel with an immobile vortex in the bulk. The measurement scheme is fully electrical and deterministic: edge vortices ($\pi$-phase domain walls) are created on demand by a voltage pulse at a Josephson junction and the braiding with a Majorana zero-mode in the bulk is detected by the charge produced upon their fusion at a second Josephson junction.
\end{abstract}
\maketitle

\textit{Introduction ---} Non-Abelian anyons have the property that a pairwise exchange operation may produce a different state, not simply related to the initial state by a phase factor \cite{Ste08}. Because such ``braiding'' operations are protected from local sources of decoherence they are in demand for the purpose of quantum computations \cite{Nay08}. Charge $e/4$ quasiparticles in the $\nu=5/2$ quantum Hall effect were the first candidates for non-Abelian statistics \cite{Moo91}, followed by vortices in topological superconductors \cite{Rea00,Iva01}. 

Because experimental evidence for non-Abelian anyons in the quantum Hall effect \cite{An11,Wil13} has remained inconclusive, the experimental effort now focuses on the superconducting realizations \cite{Lut18}. While the mathematical description of the braiding operation (the Clifford algebra) is the same in both realizations, the way in which braiding is implemented is altogether different: In the quantum Hall effect one uses the chiral motion along the edge to exchange pairs of non-Abelian anyons and demonstrate non-Abelian statistics \cite{Das05,Ste06,Bon06}. In contrast, in a superconductor the non-Abelian anyons are midgap states (``zero-modes'') bound to a defect (a vortex \cite{Vol99,Fu08} or the end-point of a nanowire \cite{Kit01,Lut10,Ore10}). Because they are immobile, existing proposals to demonstrate non-Abelian statistics do not actually exchange the zero-modes in real space \cite{Bon08,Ali11,Hec12,Vij16,Kar17}. 

Topological superconductors do have chiral edge modes \cite{Rea00}, and recent experimental progress \cite{He17} has motivated the search for ways to use the chiral motion for a braiding operation \cite{Lia17}. The obstruction one needs to overcome is that the Majorana fermions which propagate along the edge of a superconductor have conventional \textit{fermionic} exchange statistics. In the quantum Hall effect each charge $e/4$ quasiparticle contains a zero-mode and the exchange of two quasiparticles is a non-Abelian operation on a topological qubit encoded in the zero-modes. However, Majorana fermions contain no zero-mode which might encode a topological qubit, one needs vortices for that.

In this paper we show how one can exploit the chiral motion along the edge of a topological superconductor to exchange zero-modes in real space. The key innovative element of our design, which distinguishes it from Ref.\ \onlinecite{Lia17}, is the use of a biased Josephson junction to \textit{on demand} inject a pair of isolated vortices into chiral edge channels. Previous studies of such ``edge vortices'' relied on quantum fluctuations of the phase to create a vortex pair in the superconducting condensate \cite{Akh09,Nil10,Cla10,Hou11}, but here the injection is entirely \textit{deterministic}. When the two mobile edge vortices encircle a localized bulk vortex their fermion parity switches from even to odd, as a demonstration of non-Abelian braiding statistics. The entire operation, injection--braiding--detection, can be carried out fully electrically, without requiring time-dependent control over Coulomb interactions or tunnel probabilities.

\begin{figure*}[tb!]
\centerline{\includegraphics[width=0.8\linewidth]{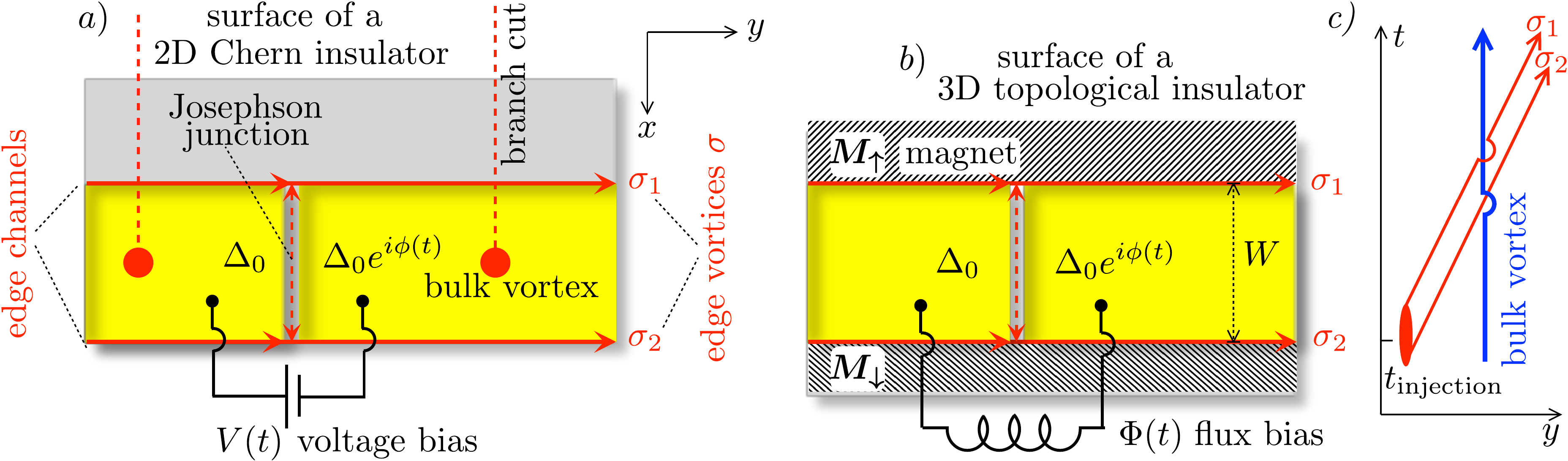}}
\caption{Panels \textit{a)} and \textit{b)}: Josephson junction geometries to deterministically inject a pair of edge vortices $\sigma_1$, $\sigma_2$ in chiral edge channels at opposite boundaries of a superconductor (yellow). The injection happens in response to a $2\pi$ increment in the superconducting phase difference $\phi(t)$, driven by a time-dependent voltage $V(t)$ or flux $\Phi(t)$. In panel \textit{a)} edge vortex $\sigma_1$ crosses the $2\pi$ branch cut of a bulk vortex, resulting in a fermion parity switch. Panel \textit{c)} shows the corresponding braiding of world lines in space-time: an overpass indicates that the vortex crosses a branch cut. 
}
\label{fig_injector}
\end{figure*}
 
\textit{Edge vortex injection ---} Fig.\ \ref{fig_injector} shows different ways in which the edge vortex can be injected: driven by a flux bias or by a voltage bias over a Josephson junction. We show two possible physical systems that support chiral edge channels moving in the \textit{same} direction on opposite boundaries of the superconductor. Both are hybrid systems, where a topologically trivial superconductor (spin-singlet \textit{s}-wave pair potential $\Delta_0$) is combined with a topologically nontrivial material: a 2D Chern insulator (quantum anomalous Hall insulator) \cite{Qi10,He17} (panel a) or a 3D topological insulator gapped on the surface by ferromagnets with opposite magnetisation $M_{\uparrow,\downarrow}$ \cite{Akh09,Fu09} (panel b).

The superconducting phase difference $\phi(t)$ across the Josephson junction is incremented with $2\pi$ by application of a voltage pulse $V(t)$ (with $\int V(t)dt=h/2e$), or by an $h/2e$ increase of the flux $\Phi(t)$ through an external superconducting loop. If the width $W$ of the superconductor is large compared to the coherence length $\xi_0=\hbar v/\Delta_0$, the edge channels at $x=\pm W/2$ are not coupled by the Josephson junction --- except when $\phi$ is near $\pi$, as follows from the junction Hamiltonian \cite{Fu08,Fu09}
\begin{equation}
H_{\rm J}=vp_x\sigma_z+\Delta_0\sigma_y\cos(\phi/2).\label{HJJ}
\end{equation}
The Pauli matrices act on excitations moving in the $\pm x$ direction with velocity $v$, in a single mode for $\xi_0$ large compared to the thickness of the junction in the $y$-direction.

At $\phi=\pi$ a Josephson vortex passes through the superconductor \cite{Pot13,Par15}. A Josephson vortex is a $2\pi$ phase winding for the pair potential, so a $\pi$ phase shift for an unpaired fermion. As explained in Ref.\ \onlinecite{Fen07}, the passage of the Josephson vortex leaves behind a pair of edge vortices: a phase boundary $\sigma(y)$ on each edge, at which the phase of the Majorana fermion wave function $\psi(y)$ jumps by $\pi$. Because of the reality constraint on $\psi$, a $\pi$ phase jump (a minus sign) is stable: it can only be removed by merging with another $\pi$ phase jump. And because the phase boundary is tied to the fermion wave function, it shares the same chiral motion, $\sigma(y,t)=\sigma(y-vt)$.

\textit{Braiding of an edge vortex with a bulk vortex ---} 
Two vortices may be in a state of odd or even fermion parity, meaning that when they fuse they may or may not leave behind an unpaired electron. The fermion parity of vortices $\sigma_1$ and $\sigma_2$ is encoded in the $\pm 1$ eigenvalue of the parity operator $P_{12}=i\gamma_1\gamma_2$, where $\gamma_n$ is the Majorana operator associated with the zero-mode in vortex $n$ \cite{note2}. The two edge vortices are created at the Josephson junction in a state of even fermion parity, $P_{12}=+1$, but as illustrated in Fig.\ \ref{fig_injector}\textit{a} that may change as they move away from the junction: If one of the edge vortices, say $\sigma_1$, crosses the branch cut of the phase winding around a bulk vortex, $\gamma_1$ picks up a minus sign and the fermion parity $P_{12}\mapsto -1$ switches from even to odd \cite{Iva01}. This is the essence of the non-Abelian braiding statistics of vortices.

\begin{figure}[tb!]
\centerline{\includegraphics[width=1\linewidth]{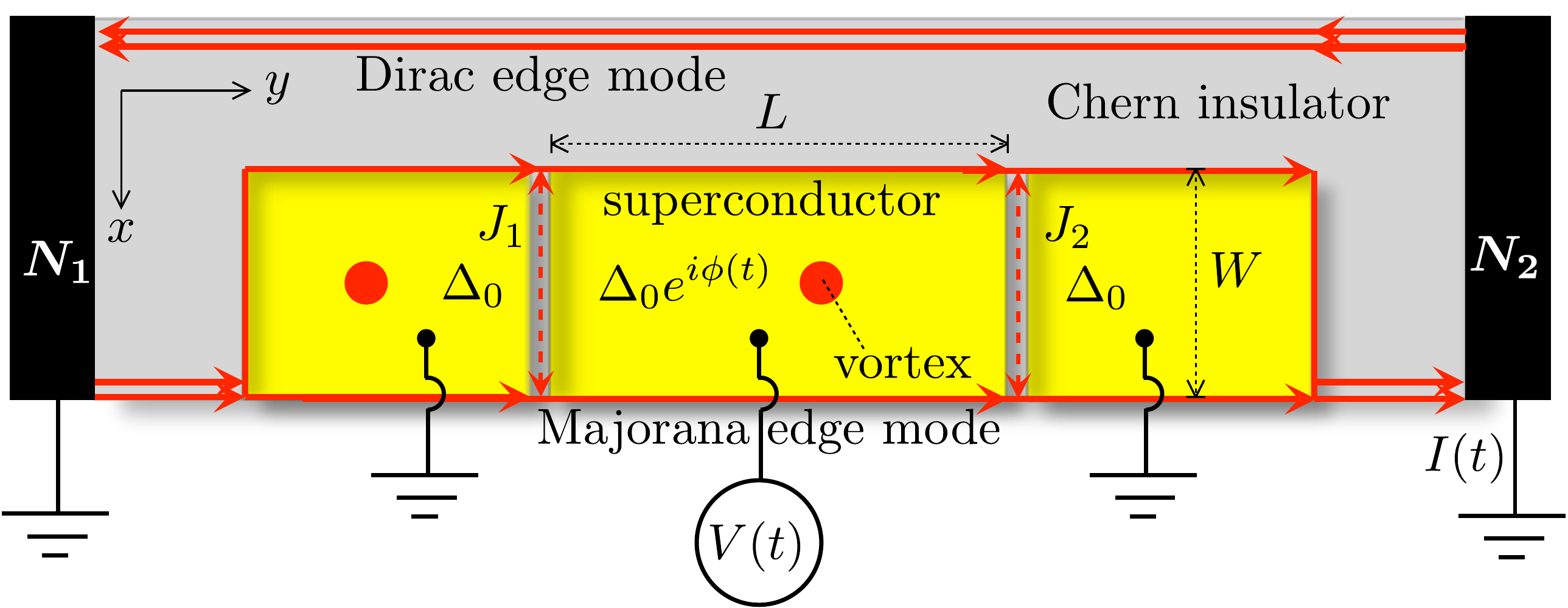}}
\caption{Starting from the layout of Fig.\ \ref{fig_injector}\textit{a}, we have inserted a second Josephson junction ($J_2$) and we have added normal metal contacts ($N_1$, $N_2$) to measure the current $I(t)$ carried by the edge modes in response to the voltage $V(t)$ applied to the superconductor. A unit charge per $2\pi$ increment of $\phi$ is transferred from the superconductor into the normal metal contact. The counterpropagating Dirac edge mode along the upper edge of the Chern insulator is decoupled from the superconductor and plays no role in the analysis.
}
\label{fig_layout}
\end{figure}

\textit{Detection of the fermion-parity switch ---}
Fig.\ \ref{fig_layout} shows the voltage-biased layout for a fully electrical measurement. The fermion parity of the edge vortices cannot be detected if they remain separated on opposite edges, so we first fuse them at a second Josephson junction. The characteristic time scale of the injection process is the time $t_{\rm inj}\simeq (\xi_0/W)(d\phi/dt)^{-1}$ when $\phi(t)$ is within $\xi_0/W$ from $\pi$, and if the distance $L$ between the two Josephson junctions is less than $vt_{\rm inj}$ we can neglect the time delay between the injection at the first junction $J_1$ and the fusion at the second junction $J_2$. This is convenient, because then the whole process can be driven by a single voltage pulse $V(t)$ applied to the region $|y|<L/2$ between the two junctions, relative to the grounded regions $y<-L/2 $ and $y>L/2$ outside.

Both these grounded regions are connected to normal metal electrodes $N_1$ and $N_2$ and the electrical current $I(t)$ between them is measured. As we will now show, the transferred charge $Q=\int I(t)dt$ is quantized at unit electron charge if the region between the Josephson junctions contains a bulk vortex, while $Q=0$ if it does not.

\textit{Mapping onto a scattering problem ---} Tunneling of edge vortices driven by quantum fluctuations of the phase is a many-body problem of some complexity \cite{Fen07}. We avoid this because we rely on an external bias to inject the edge vortices, hence the phase $\phi(t)$ can be treated as a classical variable with a given time dependence. 

The dynamics of the Majorana fermions remains fully quantum mechanical, governed by the Hamiltonian
\begin{equation}
H=i\begin{pmatrix}
-  v\partial/\partial y&-\mu[y,\phi(t)]\\
\mu[y,\phi(t)]&-  v\partial/\partial y
\end{pmatrix}\equiv vp_y\sigma_0+\mu\sigma_y.\label{Htdef}
\end{equation}
(We set $\hbar=1$.) The $2\times 2$ Hermitian matrix $H$ acts on the Majorana fermion wave functions $\Psi=(\psi_1,\psi_2)$ at opposite edges of the superconductor, both propagating in the $+y$ direction (hence the unit matrix $\sigma_0$) The interedge coupling $\mu$ multiplies the $\sigma_y$ Pauli matrix to ensure that $H$ is purely imaginary and the wave equation $\partial\Psi/\partial t=-iH\Psi$ is purely real (as it should be for a Majorana fermion).

For low-energy, long-wavelength wave packets the $y$-dependence of the interedge coupling may be replaced by a delta function, $\mu[y,\phi(t)]= v\delta(y)\eta(t)$. This ``instaneous scattering approximation'' \cite{Kee06} is valid if the transit time $t_{\rm transit}\simeq L/v$ of the wave packet through the system is short compared to the characteristic time scale $t_{\rm inj}$ of the vortex injection, hence if $d\phi/dt\ll v\xi_0/A_{\rm junction}$, where $A_{\rm junction}= WL$ is the area of the region between $J_1$ and $J_2$. In this regime there is no need to explicitly consider the vortex dynamics in between the Josephson junctions, instead we can treat this as a scattering problem ``from the outside''. 
 
Incoming and outgoing states are related by
\begin{equation}
\Psi_{\rm out}(E)=\int_{-\infty}^\infty \frac{d\omega}{2\pi}\, S(\omega)\Psi_{\rm in}(E-\omega),\label{PsiES}
\end{equation}
where $S(\omega)$ is the adiabatic (or ``frozen'') scattering matrix,
\begin{equation}
S(\omega)=\int_{-\infty}^\infty dt\,e^{i\omega t}S(t),\;\;S(t)=\exp\bigl(-i\eta(t)\sigma_y\bigr),\label{Stdef}
\end{equation}
describing the scattering at $E=0$ for a fixed $\phi(t)$.

As we shall see in a moment, the transferred charge is independent of how $\eta(t)=\eta[\phi(t)]$ is varied as a function of time, only the net increment $\delta\eta=\eta(t\rightarrow\infty)-\eta(t\rightarrow-\infty)$ matters. When there is no vortex in the region between the two Josephson junctions $J_1$ and $J_2$ there is no difference between $\phi=0$ and $\phi=2\pi$ hence $\delta\eta=0$. On the contrary, when there is a bulk vortex in this region we find \cite{appendix}
\begin{equation}
\eta=2\arccos\left(\frac{\cos(\phi/2)+\tanh\beta}{1+\cos(\phi/2)\tanh\beta}\right),\;\;\beta=\frac{W}{\xi_0}\cos\frac{\phi}{2},\label{etaphi}
\end{equation}
hence $\delta\eta=2\pi$. More generally, when there are $N_{\rm vortex}$ vortices between $J_1$ and $J_2$ the phase increment is
\begin{equation}
\delta\eta=\pi(1-(-1)^{N_{\rm vortex}}).\label{deltaetaNvortex}
\end{equation}

\begin{figure}[tb!]
\centerline{\includegraphics[width=0.8\linewidth]{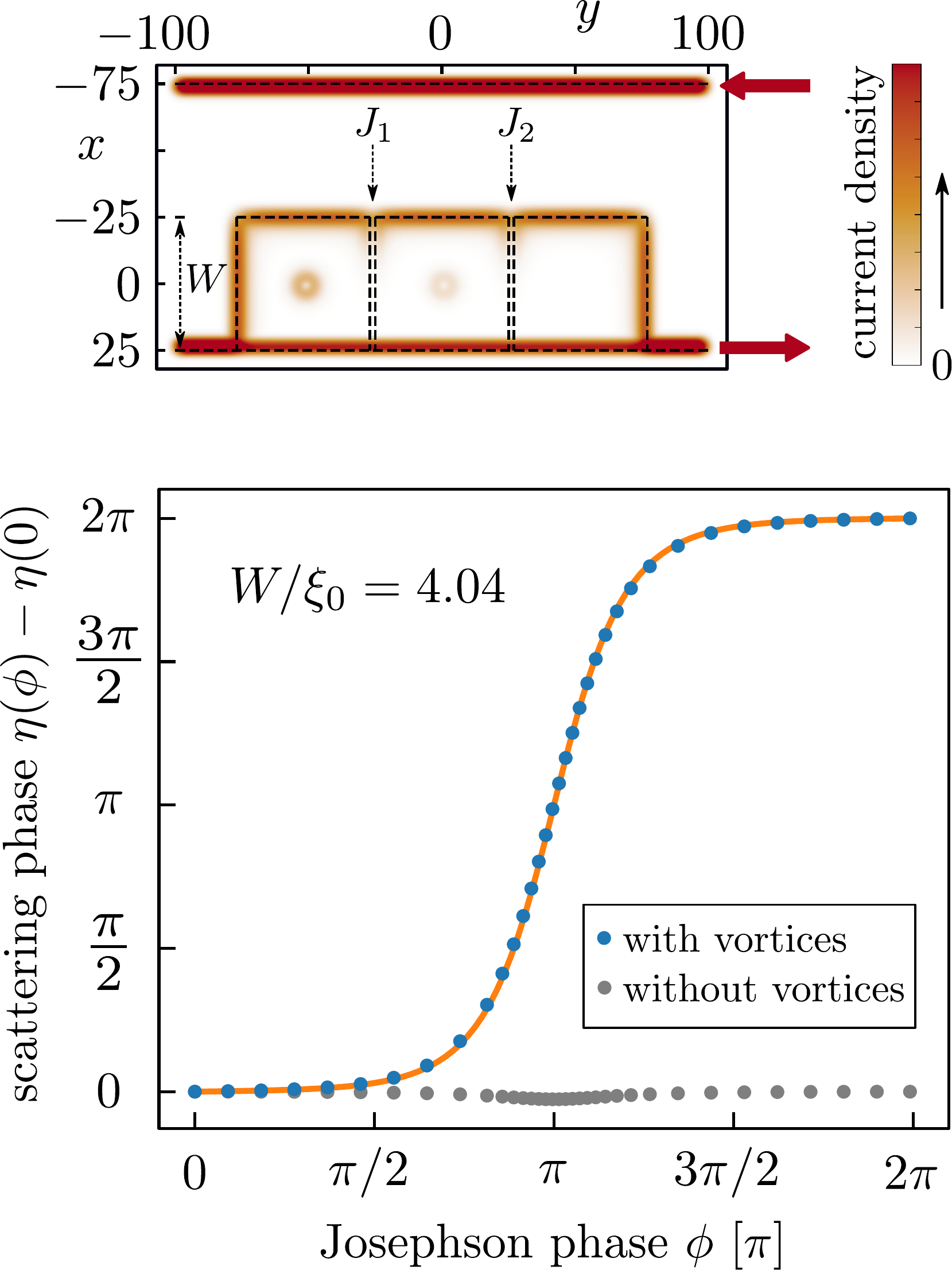}}
\caption{\textit{Bottom panel:} Scattering phase $\eta(\phi)-\eta(0)$ according to Eq.\ \eqref{etaphi} (solid curve) and as obtained numerically (blue data points) from a lattice model \cite{Qi10} of the system shown in Fig.\ \ref{fig_layout}. There are no fit parameters in the comparison, the ratio $W/\xi_0=4.04$ was obtained directly from the simulation \cite{note1}. The grey data points show the result without vortices, when there is no net increment as $\phi$ advances from 0 to $2\pi$.
\textit{Top panel:} current density in the lattice model. The two vortices are faintly visible. 
}
\label{fig_numerics}
\end{figure}

In Fig.\ \ref{fig_numerics} we show that the analytical result \eqref{etaphi} agrees well with a computer simulation (using Kwant \cite{kwant,note1}) of a lattice model of a quantum anomalous Hall insulator with induced \textit{s}-wave superconductivity \cite{Qi10}.

\textit{Transferred charge ---}
The expectation value of the transferred charge \cite{note3},
\begin{equation}
Q=e\int_{0}^\infty \frac{dE}{2\pi}\,\langle\Psi_{\rm out}^\dagger(E)\sigma_y\Psi_{\rm out}(E)\rangle,\label{Qaverage}
\end{equation}
is given at zero temperature, when 
\begin{equation}
\langle\Psi_{{\rm in},n}(E)\Psi_{{\rm in},m}(E')\rangle=\delta_{nm}\delta(E-E')\theta(-E),\label{Psiinaverage}
\end{equation}
by an integral over positive excitation energies,
\begin{equation}
Q=\frac{e}{4\pi^2}\int_{0^+}^\infty d\omega\,\omega\,{\rm Tr}\,S^\dagger(\omega)\sigma_y S(\omega).\label{QaveragezeroT}
\end{equation}
(The factor $\omega=\int_0^\infty dE\,\theta(\omega-E)$ appears from the integration over the step function.) Because $S(-\omega)=S^\ast(\omega)$ the integrand in Eq.\ \eqref{QaveragezeroT} is an even function of $\omega$ and the integral can be extended to negative $\omega$,
\begin{align}
Q&=\frac{e}{8\pi^2}\int_{-\infty}^\infty d\omega\,\omega\,{\rm Tr}\,S^\dagger(\omega)\sigma_y S(\omega)\nonumber\\
&=\frac{ie}{4\pi}\int_{-\infty}^\infty dt\,{\rm Tr}\,S^\dagger(t)\sigma_y \frac{\partial}{\partial t}S(t).\label{Brouwer}
\end{align}
This is the superconducting analogue of Brouwer's charge-pumping formula \cite{Bro98} (see Ref.\ \onlinecite{Tar15} for an alternative derivation).

Substitution of $S(t)=\exp\bigl(-i\eta(t)\sigma_y\bigr)$ results in
\begin{equation}
Q=(e/2\pi)\delta\eta=e\label{Qresult}
\end{equation}
if $N_{\rm vortex}$ is odd, while $Q=0$ if $N_{\rm vortex}$ is even.

\textit{Transferred particle number ---}
This quantized transfer of one electron charge may be accompanied by the non-quantized transfer of neutral electron-hole pairs. To assess this we calculate the expectation value of the transferred particle number, given by Eq.\ \eqref{QaveragezeroT} upon substitution of the charge operator $e\sigma_y$ by unity:
\begin{equation}
N_{\rm particles}=\frac{1}{4\pi^2}\int_{0^+}^\infty d\omega\,\omega\,{\rm Tr}\,S^\dagger(\omega) S(\omega).\label{NaveragezeroT}
\end{equation}
This integrand is an odd function of $\omega$, so we cannot easily transform it to the time domain. 

We proceed instead by calculating $S(\omega)$ from Eq.\ \eqref{Stdef}, in the approximation $\eta(t)\approx 2\arccos[-\tanh(t/t_{\rm inj})]$, accurate when $W/\xi_0\gg 1$. The result is
\begin{align}
&S(\omega)=-\frac{2\pi\omega t_{\rm inj}^2\sigma_0}{\sinh(\pi\omega t_{\rm inj}/2)}-\frac{2\pi\omega t_{\rm inj}^2\sigma_y}{\cosh(\pi\omega t_{\rm inj}/2)}-2\pi\delta(\omega)\nonumber\\
&\Rightarrow N_{\rm particles}=(84/\pi^4)\zeta(3)=1.037.\label{Nparticlesresult}
\end{align}
One can construct a special $t$-dependent phase variation \cite{note5} that makes $N_{\rm particles}$ exactly equal to unity, by analogy with the ``leviton'' \cite{Kee06,Dub13}, but even without any fine tuning the charge transfer is nearly noiseless.

\textit{Discussion ---} We have shown how the chiral motion of edge modes in a topological superconductor can be harnessed to braid a pair of non-Abelian anyons: one immobile in a bulk vortex, the other mobile in an edge vortex. The experimental layout of Fig.\ \ref{fig_layout} is directly applicable to the recently reported chiral Majorana fermion modes in quantum anomalous Hall insulator--superconductor structures \cite{He17,She18}.

While the presence of a bulk vortex and the crossing of its branch cut is essential for the charge transfer, it is of the essence for braiding that no tunnel coupling or Coulomb coupling to the edge vortices is needed. This distinguishes the braiding experiment proposed here to tunnel probes of Majorana zero-modes that can also produce a quantized charge transfer \cite{Tar15}. In the quantum Hall effect attempts to use edge modes for braiding \cite{Wil13} have been inconclusive because of Coulomb coupling with bulk quasiparticles \cite{Key15}. The superconductor offers a large gap, to suppress tunnel coupling, and a large capacitance, to suppress Coulomb coupling, which could make the edge mode approach to braiding a viable alternative to existing approaches using zero-modes bound to superconducting nanowires \cite{Bon08,Ali11,Hec12,Kar17,Vij16}. 

In the quantum Hall effect there is a drive to use quasiparticles in edge modes as ``flying qubits'' for quantum information processing \cite{Boc14}. Edge vortices in a topological superconductor could play the same role for topological quantum computation. The pair of edge vortices in the geometry of Fig.\ \ref{fig_injector}\textit{a} carries a topologically protected qubit encoded in the fermion parity. The deterministic voltage-driven injection of edge vortices that we have proposed here could become a key building block for such applications. 

\textit{Acknowledgements ---}
We have benefited from discussions with N. V. Gnezdilov. This research was supported by the Netherlands Organization for Scientific Research (NWO/OCW)  and by the European Research Council (ERC).

\clearpage

\appendix

\section{Calculation of the scattering phase shift}
\label{scatteringphase_app}

\begin{figure}[b!]
\centerline{\includegraphics[width=0.8\linewidth]{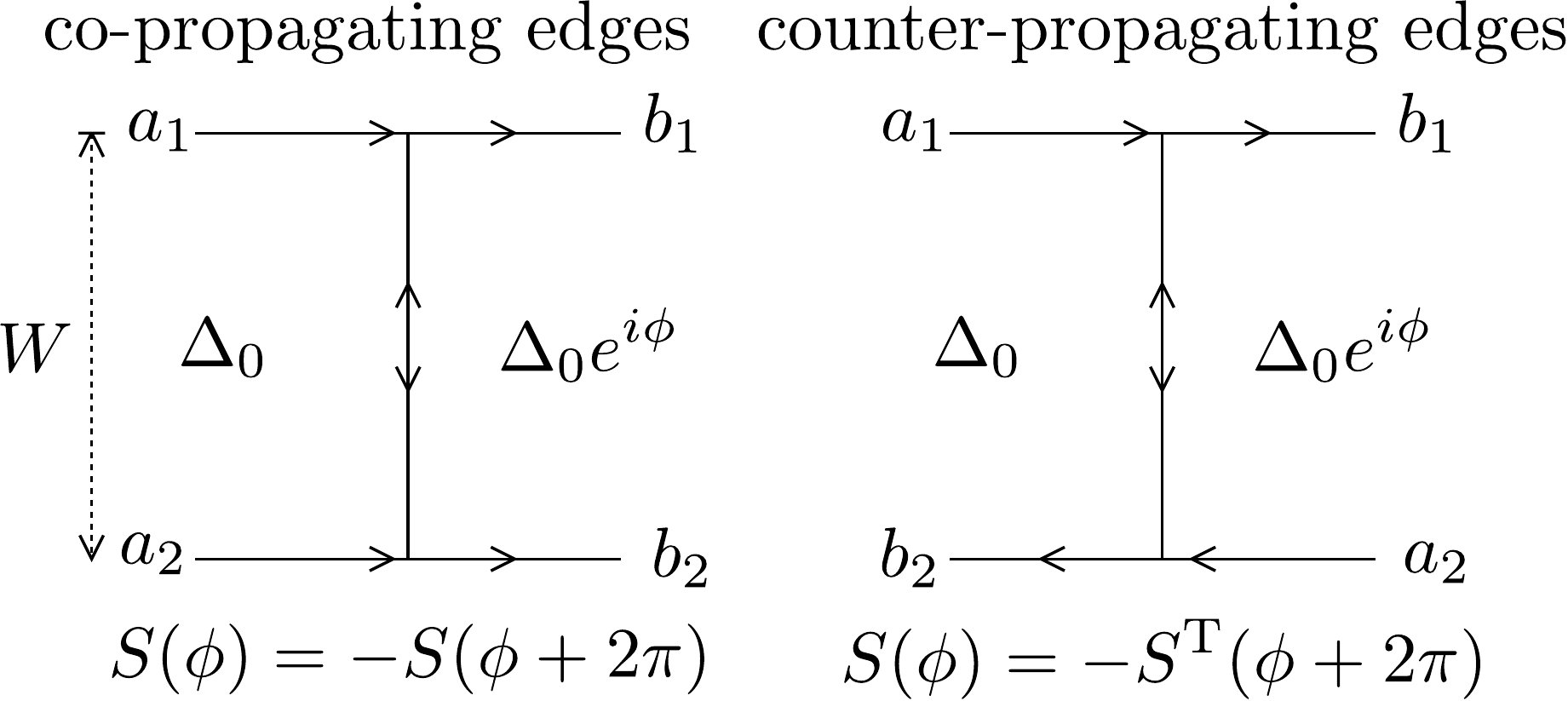}}
\caption{Two scattering geometries of a Josephson junction connecting chiral Majorana edge modes. We contrast the case of co-propagating modes in the left panel, with the case of counter-propagating modes in the right panel.}
\label{fig_JJ1}
\end{figure}

We calculate the scattering phase shift at the Fermi level in the double Josephson junction geometry of Fig.\ \ref{fig_layout}. We first consider a single Josephson junction, shown schematically in Fig.\ \ref{fig_JJ1}. We specify the phase difference $\phi$ in the interval $(-2\pi,2\pi)$, to accomodate the $4\pi$-periodicity of the junction Hamiltonian \eqref{HJJ}.

\subsection{Single Josephson junction}
\label{sec_single}

The scattering matrix $S_{\rm J}$ of the Josephson junction relates incoming and outgoing amplitudes via
\begin{equation}
S_{\rm J}\begin{pmatrix}
a_1\\
a_{2}
\end{pmatrix}=\begin{pmatrix}
b_1\\
b_2
\end{pmatrix}.\label{Sinoutrelation}
\end{equation}
At the Fermi level $S_{\rm J}\in{\rm SO}(2)$ is a $2\times 2$ orthogonal matrix with determinant $+1$, of the general form
\begin{equation}
S_{\rm J}=\begin{pmatrix}
\cos\alpha&\sin\alpha\\
-\sin\alpha&\cos\alpha
\end{pmatrix}=e^{i\alpha\sigma_y}.\label{Setadef}
\end{equation}
We seek the $\phi$-dependence of the phase shift $\alpha(\phi)$, in particular the increment $\delta\alpha=\alpha(2\pi)-\alpha(0)$.

In our geometry of co-propagating edge modes (left panel in Fig.\ \ref{fig_JJ1}), the incoming modes are on one side of the junction and the outgoing modes are at the other side. Fu and Kane \cite{Fu09} studied a different geometry with counter-propagating modes (right panel), where the two incoming modes, as well as the two outgoing modes, are on opposite sides of the junction. As we shall see, the difference is crucial for the quantization of $\delta\alpha$.

\subsubsection{Counter-propagating edge modes}
\label{sec_counter}

For counter-propagating edge modes the ${\rm SO}(2)$ scattering matrix is \cite{Fu09}
\begin{equation}
S_{\rm J}=\begin{pmatrix}
\tanh\beta&1/\cosh\beta\\
-1/\cosh\beta&\tanh\beta
\end{pmatrix},\;\;\beta=\frac{W}{\xi_0}\cos(\phi/2).\label{SJdef}
\end{equation}
If the superconducting phase difference $\phi$ across the junction is advanced by $2\pi$, a fermion crossing the junction experiences a phase shift of $\pi$. Hence the diagonal matrix elements of $S_{\rm J}$ change sign, while the off-diagonal elements do not change sign, as expressed by the symmetry relation
\begin{equation}
S_{\rm J}(\phi+2\pi)=-S^{\rm T}_{\rm J}(\phi).\label{Scountersymmetry}
\end{equation}

\begin{figure}[tb!]
\centerline{\includegraphics[width=0.8\linewidth]{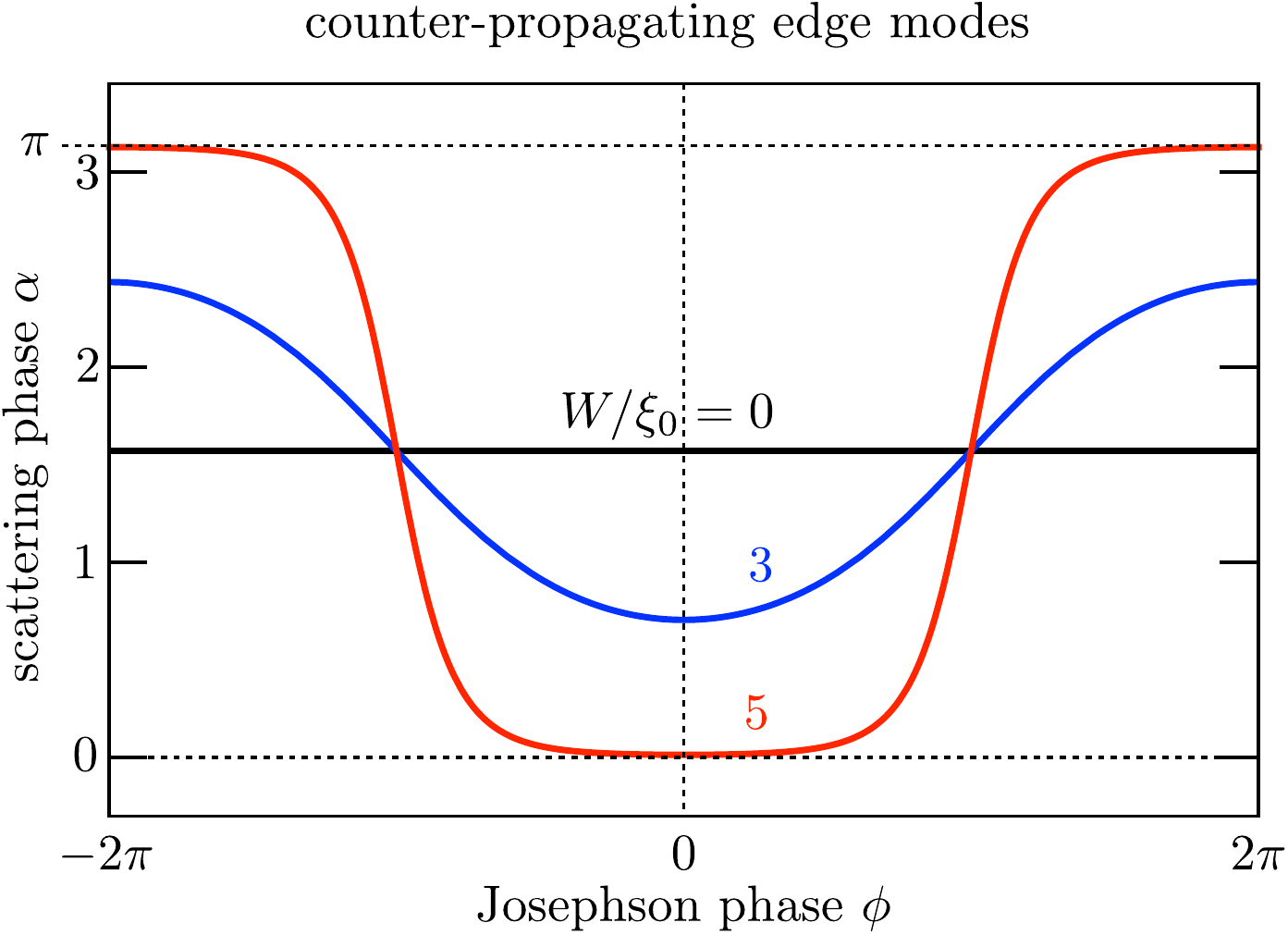}}
\caption{Plot of the $\phi$-dependence of the scattering phase shift $\alpha$ of a Josephson junction between counter-propagating Majorana edge modes. The plot is calculated from Eq.\ \eqref{alphacounter} for three values of the ratio $W/\xi_0$.}
\label{fig_alpha1}
\end{figure}

The scattering phase shift $\alpha$ in $S_{\rm J}=e^{i\alpha\sigma_y}$ from Eq.\ \eqref{SJdef} equals
\begin{equation}
\alpha={\rm arccos}\,(\tanh\beta)\in(0,\pi),\label{alphacounter}
\end{equation}
plotted in Fig.\ \ref{fig_alpha1}. The increment
\begin{equation}
\delta\alpha\approx\pi-4e^{-W/\xi_0}\label{deltaalphacounter}
\end{equation}
approaches $\pi$ for $W/\xi_0\rightarrow\infty$, but it is not quantized. Also note that $\alpha(-\phi)=\alpha(\phi)$, so the net phase increment over a $4\pi$ period is zero. Both these results for counter-propagating modes change when we consider co-propagating modes.

\subsubsection{Co-propagating edge modes}
\label{sec_co}

\begin{figure}[b!]
\centerline{\includegraphics[width=0.9\linewidth]{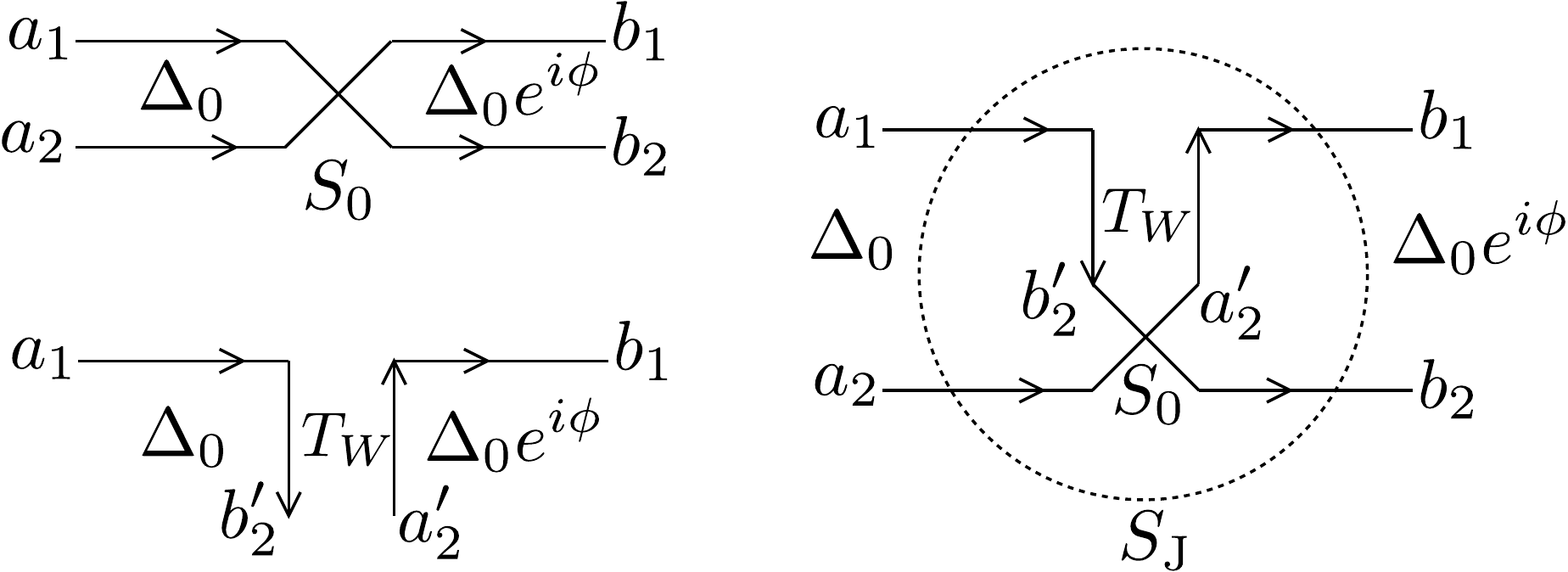}}
\caption{Scattering geometries that define $S_0$, $T_W$, and $S_{\rm J}$.}
\label{fig_JJ2}
\end{figure}

In the case of co-propagating edge modes, as in the left panel of Fig.\ \ref{fig_JJ1}, each element of the scattering matrix $S_{\rm J}$ relates amplitudes on opposite sides of the Josephson junction, so it should change sign when $\phi$ is advanced by $2\pi$. Instead of Eq.\ \eqref{Scountersymmetry} we thus have
\begin{equation}
S_{\rm J}(\phi+2\pi)=-S_{\rm J}(\phi).\label{Scosymmetry}
\end{equation}
It follows that $\alpha(2\pi)=\alpha(0)+\pi$, modulo $2\pi$, hence the phase increment
\begin{equation}
\delta\alpha=\pi\label{deltaalphaco}
\end{equation}
is \textit{exactly} quantized, independent of the ratio $W/\xi_0$. The step profile $\alpha(\phi)$ does depend on this ratio, as we now calculate.

We first consider the $W\rightarrow 0$ limit, when the Josephson junction is a point contact as in Fig. \ref{fig_JJ2}, upper left panel. The two incoming Majorana operators $\gamma_1,\gamma_2$ form an electron operator $c=(\gamma_1-i\gamma_2)/\sqrt 2$ that is transmitted through the junction with a $\phi/2$ phase shift,
\begin{align}
c_{\rm out}=e^{i\phi/2}c_{\rm in}.\label{coutcin}
\end{align}
The corresponding scattering matrix for the Majorana modes is
\begin{equation}
S_0=e^{i(\phi/2)\sigma_y}.\label{Scross}
\end{equation}

For a finite width $W$ we insert a line junction described by the Hamiltonian \eqref{HJJ}. The corresponding transfer matrix (lower left panel in Fig.\ \ref{fig_JJ2}) is
\begin{equation}
T_{\rm W}=e^{-\beta\sigma_x},\;\;\beta=\frac{W}{\xi_0}\cos(\phi/2).\label{TWdef}
\end{equation}
Combination of $S_0$ and $T_W$ (right panel in Fig.\ \ref{fig_JJ2}) produces upon mode matching the $2\times 2$ scattering matrix $S_{\rm J}$ of the entire Josephson junction,
\begin{align}
&T_W\begin{pmatrix}
a_1\\
b_1
\end{pmatrix}=\begin{pmatrix}
b'_2\\
a'_2
\end{pmatrix},\;\;S_0
\begin{pmatrix}
b'_{2}\\
a_{2}
\end{pmatrix}=\begin{pmatrix}
a'_2\\
b_2
\end{pmatrix},\nonumber\\
&\Rightarrow
S_{\rm J}\begin{pmatrix}
a_1\\
a_{2}
\end{pmatrix}=\begin{pmatrix}
b_1\\
b_2
\end{pmatrix}.
\end{align}

\begin{figure}[tb!]
\centerline{\includegraphics[width=0.8\linewidth]{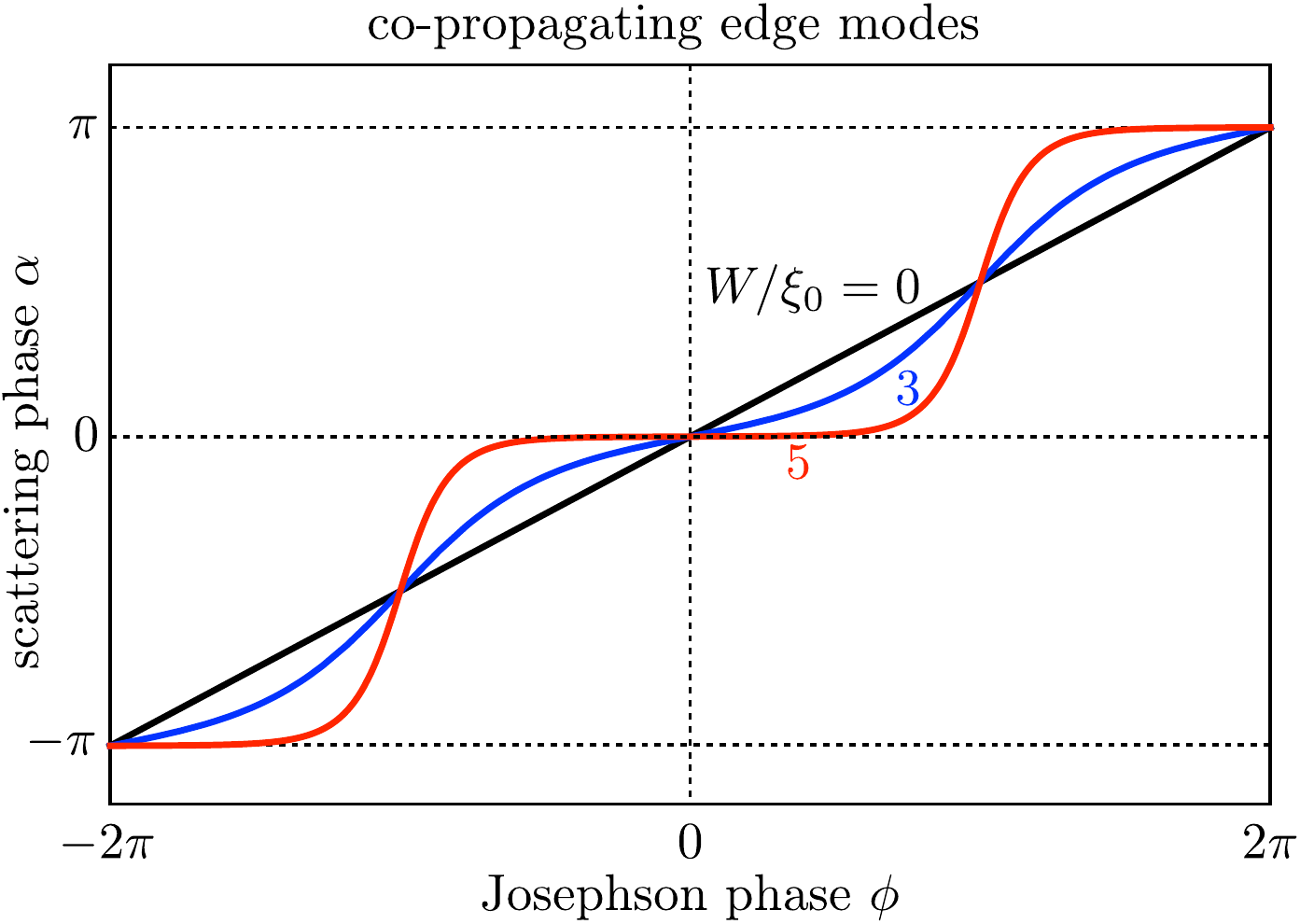}}
\caption{Same as Fig.\ \ref{fig_alpha1}, but for \textit{co}-propagating Majorana edge modes, calculated from Eq.\ \eqref{etaresult} .}
\label{fig_alpha2}
\end{figure}

The result is
\begin{widetext}
\begin{equation}
S_{\rm J}=\frac{1}{\cosh\beta+\cos(\phi/2)\sinh\beta}\begin{pmatrix}
\cos(\phi/2)\cosh\beta+\sinh\beta&\sin(\phi/2)\\
-\sin(\phi/2)&\cos(\phi/2)\cosh\beta+\sinh\beta
\end{pmatrix}.\label{SJfull}
\end{equation}
\end{widetext}
The corresponding scattering phase shift in $S_{\rm J}=e^{i\alpha\sigma_y}$ is
\begin{equation}
\alpha=\arccos\left(\frac{\cos(\phi/2)+\tanh\beta}{1+\cos(\phi/2)\tanh\beta}\right)\times{\rm sign}\,(\phi).\label{etaresult}
\end{equation}
It increases monotonically from $\alpha=-\pi$ at $\phi=-2\pi$ through $\alpha=0$ at $\phi=0$ to $\alpha=\pi$ at $\phi=2\pi$. As shown in Fig.\ \ref{fig_alpha2}, the increase starts out linearly for $W/\xi_0\ll 1$, and then becomes more and more step-function like with increasing $W$.

For $W/\xi_0\gg 1$ the $\phi$-dependence of $\alpha$ is described with exponential accuracy by
\begin{equation}
\alpha\approx\arccos(\tanh\beta)\times{\rm sign}\,(\phi),\label{alphaapprox}
\end{equation}
as in Eq.\ \eqref{alphacounter}, but now antisymmetric in $\phi$. As a consequence the net phase increment over a $4\pi$ period equals $2\pi$ rather than zero.

\subsection{Double Josephson junction}
\label{sec_double}

We combine two Josephson junctions in series with co-propagating edge modes, as in Fig.\ \ref{fig_layout}. We denote the result \eqref{SJfull} by
\begin{equation}
S_{\rm J}= S_1=e^{i\alpha_1\sigma_y},\label{S1def}
\end{equation}
with a subscript $1$ to indicate that this is the scattering matrix of the first Josephson junction ($J_1$ in Fig.\ \ref{fig_layout}).

If the second Josephson junction $J_2$ would be identical to the first, its scattering matrix would be $S_2(\phi)=S_1(-\phi)=S_1^{-1}$. We allow for a difference in the ratio $W/\xi_0$ at the two junctions, so more generally
\begin{equation}
S_2=e^{-i\alpha_2\sigma_y}.\label{S2def}
\end{equation}
The parameter $\alpha_2$ still increases by $\delta\alpha_2=\pi$ for each $2\pi$ increment of $\phi$, but it may do so with a different $\phi$-dependence than $\alpha_1$.

If there are no bulk vortices in the superconductor, the scattering matrix of the two junctions in series is simply the product $S_2 S_1$. However, if the geometry is as in Fig.\ \ref{fig_layout}, with a pair of bulk vortices on opposite sides of the first Josephson junction, we have to insert a $\sigma_z$ to account for each crossing of a branch cut, so the full scattering matrix is $S_2\sigma_z S_1 \sigma_z$. The two cases can be combined as
\begin{equation}
S_\pm =e^{-i\eta_\pm\sigma_y},\;\;\eta_\pm=\alpha_2\pm\alpha_1,\label{Spmdef}
\end{equation}
where $S_+$ and $S_-$ refer, respectively, to the situation with or without the bulk vortices. More generally, if the region between Josephson junctions $J_1$ and $J_2$ has $N_{\rm vortex}$ bulk vortices, $S_+$ applies if $N_{\rm vortex}$ is odd while $S_-$ applies if $N_{\rm vortex}$ is even.

\begin{figure}[tb!]
\centerline{\includegraphics[width=0.8\linewidth]{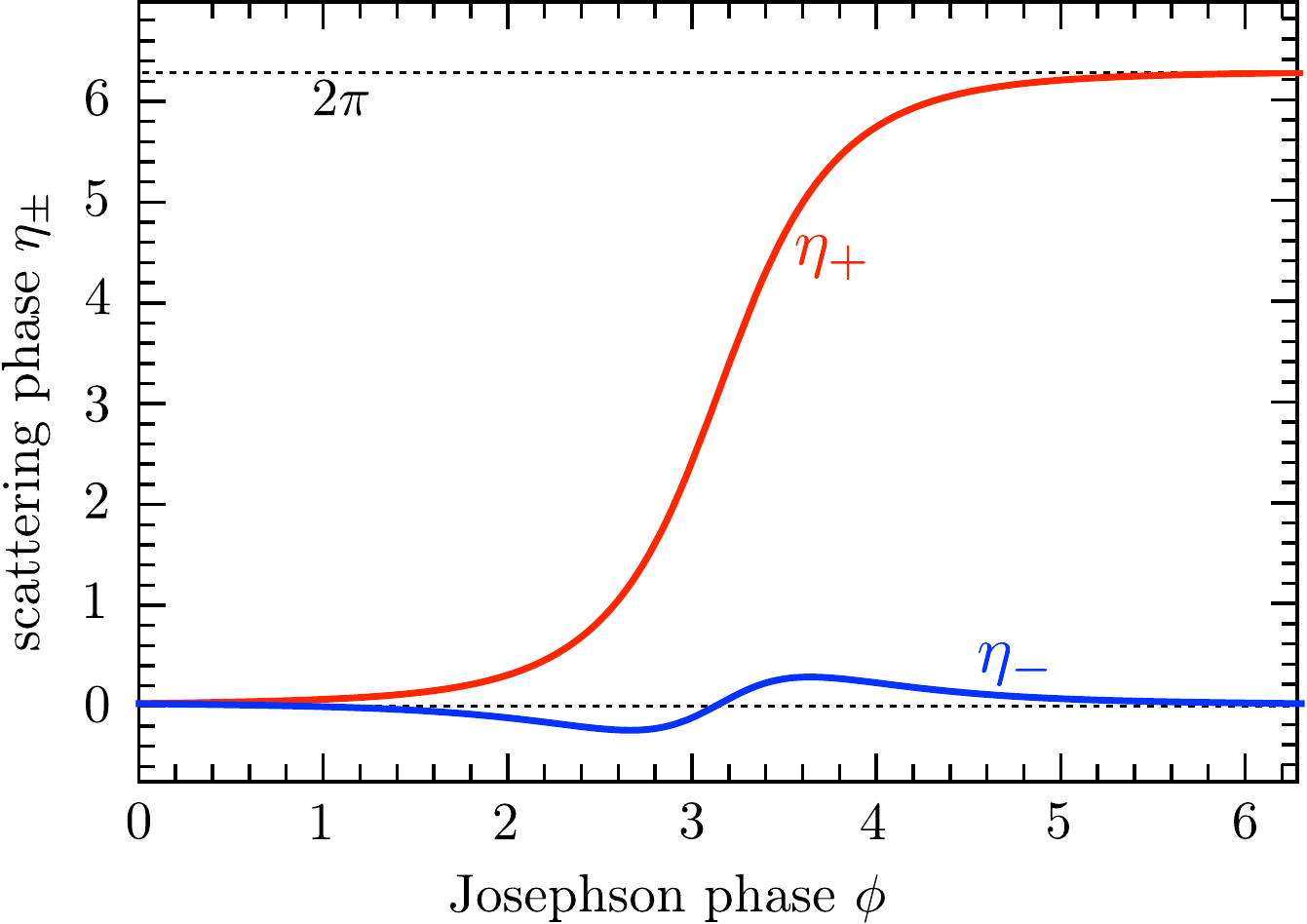}}
\caption{Plot of the $\phi$-dependence of the two scattering phase shifts $\eta_+$ (with a bulk vortex in the region between the two Josephson junctions $J_1$ and $J_2$) and $\eta_-$ (without a bulk vortex). The plot is calculated from Eqs.\ \eqref{etaresult} and \eqref{Spmdef} for $W/\xi_0=3$ in the first junction and $W/\xi_0=5$ in the second jnction.}
\label{fig_eta}
\end{figure}

When the phase $\phi$ across the Josephson junction varies from $0$ to $2\pi$ both $\alpha_1$ and $\alpha_2$ advance from $0$ to $\pi$. It follows that a $2\pi$ increment of $\phi$ induces a $2\pi$ increase of the scattering phase $\eta_+$, while $\eta_-$ has no net increase. Fig.\ \ref{fig_eta} illustrates the difference for a particular choice of parameters.

\section{Details of the numerical simulation}
\label{simulation_app}

\begin{figure}[tb!]
\centerline{\includegraphics[width=1\linewidth]{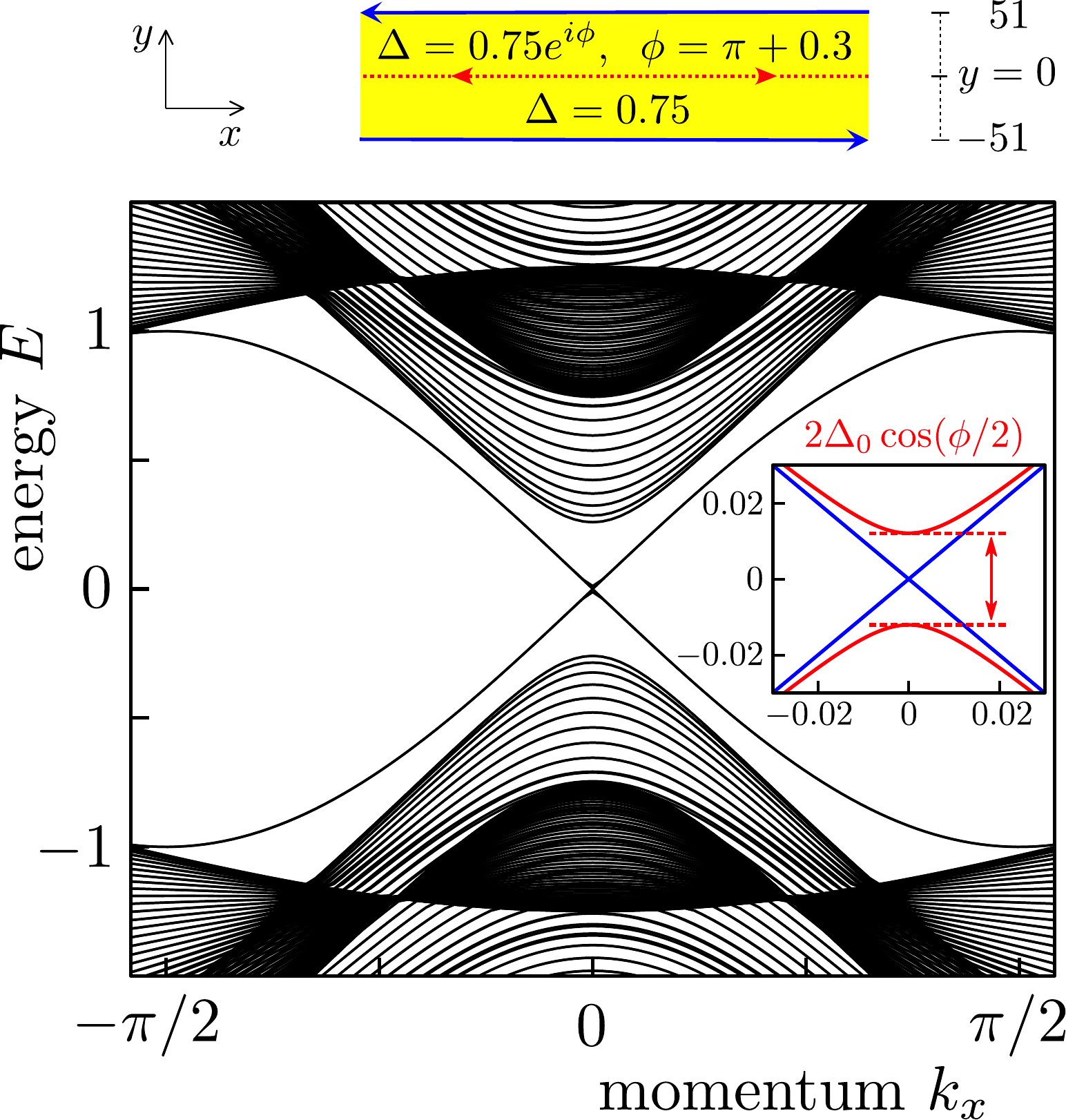}}
\caption{Dispersion relation of a superconducting strip in the region $1 < |y|<51$, with a line junction along $y=0$ ($\Delta = 0$, two lattice sites wide), separating regions with phase $0$ and phase $\phi=\pi+0.3$. Inset: Enlargement of the region near $k_x=0$, $E=0$. The blue modes are gapless chiral Majorana edge modes at the boundary of the superconductor, the red mode is a nonchiral Majorana mode in the line junction, with a gap of $2\Delta_0\cos(\phi/2)$. The effective pair potential $\Delta_0=0.0808$ is much smaller than the bare value $0.75$. }
\label{fig_dispersion}
\end{figure}

For the numerical simulation shown in Fig.\ \ref{fig_numerics} we applied the Kwant tight-binding code \cite{kwant} to the Bogoliubov-De Gennes Hamiltonian of Qi, Hughes, and Zhang \cite{Qi10}:
\begin{subequations}
\begin{align}
  {\cal H} &= 
\begin{pmatrix}
h_{0}(\bm p) - E_{\rm F} & i \Delta(\bm{r}) \tau_y \\
-i \Delta^*(\bm{r}) \tau_y & -h_{0}^*(-\bm p) + E_{\rm F}
\end{pmatrix},\\
  h_{0}(\bm p) &= (C+Bp_x^2+Bp_y^2) \tau_z + A p_x \tau_x + A p_y \tau_y.
\end{align}
\end{subequations}
The blocks of ${\cal H}$ refer to the electron-hole degree of freedom, while the Pauli matrices $\tau_\alpha$ act on the spin degree of freedom. This Hamiltonian was discretized on a two-dimensional square lattice. Lengths are measured in units of the lattice constant $a=1$ and energies in units of the hopping matrix element $t_0=1$. We also set $\hbar=1$, so that all parameters are dimensionless. 

The electron block $h_0(\bm{p})$ describes a quantum anomalous Hall insulator. We took the parameters $A=1$, $B=0.5$, $C=-0.5$, $E_{\rm F}=0$, when $h_0$ has Chern number $1$. The insulator covers the region $-75<x<25$. The bulk is gapped while the edges support a single chiral edge mode. The mode at the $x=25$ boundary moves in the $+y$ direction and a counterpropagating edge mode flows at the $x=-75$ boundary. (See Fig.\ \ref{fig_numerics}, top panel, for the geometry.)

The pair potential $\Delta$ induces spin-singlet \textit{s}-wave superconductivity in a strip $-25< x< 25$ (so $W=50$) streching from $y=-77$ to $y=77$. We inserted two line junctions $J_1,J_2$ (each two lattice sites wide), separating three superconducting islands $I_1,I_2,I_3$, by means of the profile
\begin{equation}
\Delta=\begin{cases}
0.75&{\rm if}\;\;-77 < y< -27\;\;(I_1),\\
0&{\rm if}\;\;-27<y<-25\;\;(J_1),\\
0.75\,e^{i\phi}&{\rm if}\;\;-25< y< 25\;\;(I_2),\\
0&{\rm if}\;\;25<y<27\;\;(J_2),\\
0.75&{\rm if}\;\;27< y < 77\;\;(I_3).
\end{cases}
\end{equation}
The effective gap $\Delta_0\cos(\phi/2)$ in the Josephson junction was obtained directly from the excitation spectrum. (See Fig.\ \ref{fig_dispersion}.) We found $\Delta_0=0.0808$ --- much smaller than the bare gap of 0.75. At $\phi=\pi$ the gap closes, producing a linear dispersion along the Josephson junction with velocity $v=1$ --- the same as the velocity of the edge modes. The corresponding coherence length is $\xi_0=v/\Delta_0=12.38$, resulting in a ratio $W/\xi_0=4.04$. 

A pair of vortices is inserted at positions $\bm{r}_1 = (1, -51)$ in $I_1$ and $\bm{r}_2 = (1, 1)$ in $I_2$. The vortex core does not coincide with a lattice point (which are at half-integer $x,y$), so we can keep a constant ampitude $|\Delta|=0.75$ of the pair potential. Multiplication of $\Delta(\bm{r})$ by the function
\begin{equation}
f(\bm{r})= \frac{z - z_1}{|z - z_1|} \frac{|z - z_2|}{z - z_2},\;\;z=x+iy,
\end{equation}
ensures that the phase of the pair potential winds by $\pm 2\pi$ around each vortex.

The scattering matrix was calculated at an energy $E=0.001$ that is slightly offset from the Fermi level at $E=0$ to avoid the zero-mode resonance in the vortex cores. The representation $S=e^{-i\eta\sigma_y}$ in the Majorana basis corresponds to $S=e^{-i\eta\sigma_z}$ in the electron-hole basis, so the scattering phase shift $\eta$ can be calculated by comparing incident and transmitted electron wave functions along $x=25$. The edge at $x=-75$ is decoupled from the superconductor and does not contribute to $\eta$. 

At $\phi=0$ we find $\eta\neq 0$, presumably because of additional phase shifts acquired when the Dirac mode splits into two Majorana modes and back at $\bm{r}=(25,-77)$ and $\bm{r}=(25,77)$. In Fig.\ \ref{fig_numerics} we have plotted the phase increment $\eta(\phi)-\eta(0)$, to eliminate this $\phi$-independent offset. 


\begin{thebibliography}{99}
\bibitem{appendix} The calculation of the scattering phase shift $\eta(\phi)$ is given in App.\ \ref{scatteringphase_app}. Eq.\ \eqref{etaphi} for $0\leq\phi\leq 2\pi$ repeats periodically modulo $2\pi$.
\bibitem{note1} Details of the numerical simulation are given in App.\ \ref{simulation_app}.
\end{thebibliography}
\end{document}